\documentclass[%
 reprint,
nofootinbib,
 amsmath,amssymb,
 aps,
 prd,
]{revtex4-2}
\usepackage{xcolor}% add hypertext capabilities
\usepackage{hhline}
\usepackage[usenames,dvipsnames,svgnames]{xcolor}
\usepackage[colorlinks=true,citecolor=purple,urlcolor=blue,linkcolor=olive]{hyperref}
\usepackage{graphicx}% Include figure files
\usepackage{dcolumn}% Align table columns on decimal point
\usepackage{bm}% bold math
\usepackage{ulem}
\usepackage{array}
\usepackage{hyperref}% add hypertext capabilities
\usepackage{subfigure}
\usepackage{tabularray}
\UseTblrLibrary{booktabs}

\usepackage{multirow}
\usepackage{natbib}
\usepackage{mathrsfs}
\begin{abstract}

We investigate, for the first time, the non-radial $f$-mode oscillations of neutron stars (NSs) with a dark energy (DE) core utilising a fully general relativistic treatment. We construct the stellar profile by considering a  relativistic mean field model for nuclear matter and a modified Chaplygin fluid like prescription for DE. 
We compute the stellar structure, $f$-mode oscillations, tidal deformability, and gravitational wave (GW) energy and strain of the NS with DE core by varying the DE equation of state (EoS) parameters.
Our study reveals that the inclusion of DE softens EoS and reduces the maximum mass compared to the pure neutron star. The obtained mass-radius values are consistent with observational constraints from massive pulsars such as PSR J0030+0451 and PSR J0740+6620. 
We find that the extent of the DE core, determined by the transition density, plays a crucial role in modifying the stellar structure and oscillation properties, with lower transition densities producing appreciable changes over a wider mass range. The $f$-mode frequencies and damping times exhibit systematic modifications, for higher mass configurations, while the correlation between the $f$-mode frequency and tidal deformability remains consistent with observational constraints from GW170817 and GW190814. Further, the normalised oscillation energy distribution remains nearly universal, with only minimal deviations for the higher mass configurations. Finally, we estimate the characteristic GW strain associated with the $f$-mode oscillations and find that the predicted signals lie within the sensitivity band of future third generation GW detectors.
\end{abstract}

\begin{document}

\title{Gravitational Imprints of Dark Energy in Neutron Stars}

 \author{O. P. Jyothilakshmi}
 \email{op\_jyothilakshmi@cb.students.amrita.edu}
 \author{Lakshmi J. Naik}
 \email{jn\_lakshmi@cb.students.amrita.edu}
 \author{V. Sreekanth}
 \email{v\_sreekanth@cb.amrita.edu}

\affiliation{Department of Physics, Amrita School of Physical Sciences, Amrita Vishwa Vidyapeetham, Coimbatore, India}

\date{\today}
\maketitle
%%%%%%%%%%%%%%%%%%%%%%%%%%%%%%%%%%%%%%%%%%%%%%%%%%%%%%%%%%%%%%%%%%%%%%%%%%%
\section{Introduction}
For more than half a century, neutron stars (NSs) have remained one of the most fascinating and extensively studied astrophysical objects. They are highly dense compact objects that serve as natural laboratories for studying matter under extreme conditions. The interior composition  of a NS is largely unknown, particularly at the high density region, where exotic matter may arise. Several forms of exotic matter, such as hyperons, deconfined quark matter, Bose-Einstein condensates, dark matter have been proposed to appear in the high density interior cores of NSs. The presence of them is found to modify the structural and dynamic properties of compact stars~\cite{Glendenning:1997wn,LATTIMER2006479,Lattimer:2021emm}. The detection of gravitational waves (GWs) from compact binary mergers by the LIGO-Virgo-KAGRA collaboration, along with precise pulsar measurements of mass, radius, and tidal deformability has recently opened a new avenue for constraining the equations of state (EoSs) through multi-messenger astronomy~\cite{LIGOScientific:2017vwq,LIGOScientific:2018cki,LIGOScientific:2020zkf}.
 
The detection of GWs produced from the excitations of quasi-normal modes is expected to play a crucial role in the era of NS asteroseismology. The oscillation and frequency measurements of these modes carry valuable information about the internal structure and composition of compact objects, which thereby helps to constrain the EoS of dense matter~\cite{Andersson:2021qdq}. This motivates the study of GW driven non-radial oscillations of various stellar objects. 
The non-radial oscillations of general relativistic models were first studied in detail by Thorne and Campolattaro in 1967~\cite{1967ApJ...149..591T}, by considering linear perturbations in a static, spherically symmetric stellar background. They developed a fifth order system of ordinary differential equations governing the fluid and metric perturbations. Subsequent studies extended this framework to develop a deeper understanding of the relativistic oscillation spectrum and associated GW emission~\cite{Thorne1969}.   
Later, Lindblom and Detweiler simplified this formalism to a fourth order system of equations, giving precise understanding of different modes of oscillations~\cite{Lindblom:1983ps,Detweiler:1985zz}. 
This fully relativistic perturbative formalism has become a standard approach for studying the non-radial oscillation spectrum of various stellar configurations and subsequent establishment of precise empirical relations between the oscillation spectra and other stellar properties~\cite{Sotani:2001bb,Kunjipurayil:2022zah,Karkevandi:2021ygv,Das:2021dru,Shirke:2024ymc,Thakur:2024btu}. 
Among various non-radial oscillation modes the fundamental oscillation ($f$-) modes are of great interest, in view of their sensitivity to stellar properties and strong coupling to gravitational radiations. Moreover, the frequency of $f$-modes lies in the range that can be measured by future GW detectors. The $f$-mode oscillations are widely studied within various stellar models~\cite{Kunjipurayil:2022zah,Shirke:2024ymc,Jyothilakshmi:2024zqn,Jyothilakshmi:2025wru,Jyothilakshmi:2024xtl,Thakur:2024btu,Kumar:2025cro}. 

The dimensionless tidal deformability $\Lambda$ measures the extent to which a NS is tidally deformed by its companion during the inspiral stage of binary merger. This quantity is extremely sensitive to the stellar radius $R$ and compactness ($M/R$)~\cite{Hinderer2008}. 
Motivated by the pioneering work of Thorne on GW emission from stellar oscillations~\cite{Thorne1969}, many studies have investigated GW energy from NS oscillations, the associated damping mechanisms and detectability of emitted signals~\cite{Benhar:2004xg}. These studies have evolved from early analyses of relativistic stellar pulsations to modern investigations of $f$-mode asteroseismology and GW emission from compact stars with exotic interiors~\cite{Ball:2023nme,Shirke:2024ymc,Zheng:2025xlr}. 

More recently, various stellar properties of NSs with dark energy (DE) like matter in the inner core have attracted attention~\cite{Pretel:2024tjw,Pretel:2024vvt}. 
Chapline was the first to propose the idea of compact stars made of DE fluid~\cite{Chapline:2004jfp}. 
Among the various phenomenological descriptions of DE stars, the Chaplygin gas EoS has been widely employed due to its ability to reproduce DE like behavior through a negative pressure fluid prescription~\cite{Kamenshchik:2001cp,Kahya:2015dpa}. The stellar properties of pure DE stars (DESs), including their structure, tidal deformability, and both radial and non-radial oscillations have been studied extensively over the past few years~\cite{Panotopoulos:2020kgl,Panotopoulos:2021dtu,Pretel:2023nhf,Jyothilakshmi:2024zqn,Jyothilakshmi:2024zso,Jyothilakshmi:2025odz}. 
The radial oscillations of NSs containing a DE core have recently been investigated in Ref.~\cite{Pretel:2024vvt}. However, the non-radial oscillation spectrum of such hybrid configurations, which can provide direct GW signatures of the DE component, have not yet been explored. 
Motivated by this gap, we investigate the global stellar properties and non-radial oscillation modes of NSs containing a DE core, in this work. The presence of DE matter is expected to modify both the equilibrium structure and dynamical response of the star, leaving imprints on observable quantities such as the mass, radius, tidal deformability, and GW characteristics of the oscillation modes. These observables provide complementary probes of the internal structure and offer a means to assess the viability of hybrid NS models containing DE matter in light of constraints from pulsar measurements, X-ray observations, and GW detections. We therefore study the $f$-mode spectrum, damping times, oscillation energy, tidal deformability, and associated GW observables within a fully relativistic framework, and examine their dependence on the DE EoS parameters and the extent of the DE core.
 
The paper is organised as follows. In Section~\ref{sec:eos}, we present the construction of EoS used for NSs with DE core. Section~\ref{sec:GR} gives the review of general relativistic treatment used to determine stellar structure, $f$-mode oscillation spectra, tidal deformability and GW energy and strain. The calculated results are presented in Section~\ref{sec:results}, followed by the summary and conclusion in Section~\ref{sec:summary}. 
\textit{Notations and conventions}: 
Throughout the manuscript we set $G=c=\hbar=1$ and use the metric convention of  $(+1,\,-1,\,-1,\,-1)$. We denote M$_\odot$ as mass of the Sun. 
%%%%%%%%%%%%%%%%%%%%%%%%%%%%%%%%%%%%%%%%%%%%%%%%%%%%%%%%%%%%%%%%%%%%%%%%%%%
\section{Equation of state}
\label{sec:eos}

In this work, we adopt a phenomenological approach and model the stellar interior using a generalized Chaplygin fluid~\cite{Kamenshchik:2001cp} description of DE matter in the core region and a realistic NS EoS in the outer region. We develop a piecewise hybrid EoS (DENS) by constructing a regular interface where both energy density and pressure are continuous. 
The Chaplygin gas EoS for DE is given by~\cite{Kahya:2015dpa}
\begin{equation}
p = A\, \varepsilon-B/\varepsilon^\alpha;
\end{equation}
where $p$ is the pressure, $\varepsilon$ is the energy density, $A$ is a dimensionless positive constant, $B$ is a positive constant with units of meters, $m^{-2( 1 + \alpha )}$ units
 and $\alpha$ is a constant that lies in the range $(0,1)$. The Chaplygin gas EoS has been proposed as a candidate for DE due to its ability to produce accelerated expansion in cosmological models. This DE EoS has been extensively used to study various astrophysical aspects of compact objects composed of DE~\cite{Lobo:2005uf,Chan:2008ui,Yazadjiev:2011sd,Rahaman:2011hd,Beltracchi:2018ait,Panotopoulos:2021dtu,Panotopoulos:2020kgl,Pretel:2023nhf,Panotopoulos:2021dtu,Jyothilakshmi:2024zqn,Jyothilakshmi:2024zso,Jyothilakshmi:2025odz}. In this study, we explore the effects of incorporating a DE core into NSs on their structural and oscillation properties. 

%%%%%%%%%%%%%%%%%%%%%%%%%%%%%%%%%NS-model%%%%%%%%%%%%%%%%%%%%%%%%%%%%%%%%%%%%
The outer layers of the NS are described by a realistic nuclear matter EoS. 
The model considered for neutron matter here is a relativistic mean-field (RMF) model~~\cite{Serot:1979dc,Gambhir:1990uyn,Serot:1997xg} with the Lagrangian density given by~\cite{Xia:2022dvw,Niu:2025tvd}
\begin{align}
\mathcal{L}=&\sum_{i=n,\,p}
\bar{\psi}_i\left[i\gamma^\mu \partial_\mu-\gamma^0\left(g_{\omega}\omega+g_{\rho}\rho \tau_i + \mathscr{A} q_i\right)
-m_i^*
\right]
\psi_i \nonumber
\\
&+
\sum_{l=e,\,\mu}
\bar{\psi}_l
\left(
i\gamma^\mu \partial_\mu
-m_l
+e\gamma^0 \mathscr{A}
\right)
\psi_l
-\frac{1}{4}F_{\mu\nu}F^{\mu\nu}  \nonumber
\\
&+
\frac{1}{2}\partial^\mu \sigma \partial_\mu \sigma
-\frac{1}{2}m_\sigma^2 \sigma^2
-\frac{1}{4}\omega_{\mu\nu}\omega^{\mu\nu}
+\frac{1}{2}m_\omega^2 \omega^2 \nonumber
\\
&-
\frac{1}{4}\rho_{\mu\nu}\rho^{\mu\nu}
+\frac{1}{2}m_\rho^2 \rho^2
+U(\sigma,\omega,\rho)\cdot 
\label{eq:lagrangian}
\end{align}
Here, $\psi_i$ and $\psi_l$ denote the nucleonic ($n,p$) and leptonic ($e,\mu$) Dirac fields, respectively. 
The mesonic fields $\sigma$, $\omega$, and $\rho$ correspond to the scalar-isoscalar, vector-isoscalar, and vector-isovector mesons, respectively; while $\mathscr{A}$ denotes the electromagnetic field.
Now, $\tau_n=-\tau_p=1$ is the third component of isospin of nucleons, and $q_p=-q_e=-q_\mu = e$ and $q_n=0$ are the corresponding charges. The quantity $m_i^*=m_i+g_\sigma\sigma$ denotes the effective nucleonic mass and $m_l$ represents the leptonic mass. 
The second rank field tensors appearing in the Lagrangian has the general form 
%\begin{equation*}
 $\Phi_{\mu\nu} = \partial_{\mu}\Phi_{\nu} - \partial_{\nu}\Phi_{\mu}$.  
%\end{equation*}
Assuming time-reversal symmetry in the RMF 
approximation, the bosonic fields $\sigma$, $\omega$, $\rho$ and $\mathscr{A}$ take corresponding mean values and only the temporal components remain non-zero. Now, the non-vanishing components of the field tensors reduce to
\begin{equation*}
    \omega_{0i}=-\omega_{i0}=\partial_i \omega,\,\rho_{i0}=-\rho_{0i}=\partial_i\rho,\,F_{i0}=-F_{0i}=\partial_i \mathscr{A}.
\end{equation*}
The mesonic interaction potential is given as
\begin{equation}
U(\sigma,\omega,\rho) = -\frac{1}{3}g_2\sigma^3-\frac{1}{4}g_3\sigma^4 +\frac{1}{4}c_3\omega^4 +\frac{1}{4}d_3\rho^4 ;
\end{equation}
where, $g_2$, $g_3$, $c_3$, and $d_3$ are the non-linear coupling constants.

The equations of motion for bosonic fields are obtained from corresponding Euler-Lagrange equations and are given by (within RMF approximation):
\begin{align}
\left(-\nabla^2+m_\sigma^2\right)\sigma
&=
-g_\sigma n_s
-g_2\sigma^2
-g_3\sigma^3,
\\
\left(-\nabla^2+m_\omega^2\right)\omega
&=
g_\omega n_b
-c_3\omega^3,
\\
\left(-\nabla^2+m_\rho^2\right)\rho
&=
\sum_{i=n,p} g_\rho \tau_{3i} n_i
-d_3\rho^3,
\\
-\nabla^2 \mathscr{A}
&=
e\left(n_p-n_e-n_\mu\right).
\end{align}
Note that, we consider NS matter at zero temperature and employ the no-sea approximation. The scalar and vector densities are then given by
\begin{align}
n_s
&=
\sum_{i=n,p}\langle\bar{\psi}_i\psi_i\rangle
=
\sum_{i=n,p}
\frac{M^{*3}}{2\pi^2}
\,g
\!\left(\frac{\nu_i}{M^*}\right),
\\
n_i
&=
\langle\bar{\psi}_i\gamma^0\psi_i\rangle
=
\frac{\nu_i^3}{3\pi^2}.
\end{align}
Here, $\nu_i$ is the Fermi momentum and $g(x)=x\sqrt{x^2+1}-\operatorname{arcsinh}(x)$.

The particle numbers $N_i$ and the energy density $\varepsilon$ of the system can be obtained for any given density profiles $n_i (\vec{r})$. The ground state is fixed by minimizing $E$ under the constraint of fixed particle numbers. This results in a constant value for the chemical potential given by
\begin{equation}
\mu_i%(\vec{r})
=
\sqrt{\nu_i^{\,2}+m_i^{*2}}
+g_{\omega}\omega
+g_{\rho}\tau_i \rho
+q_i \mathscr{A} .
\end{equation}
Now, the energy-momentum tensor ($T_{\mu\nu}$) for the system can be obtained from the Lagrangian and comparing with the same for the perfect fluid prescription at rest, we obtain the energy density of the system as 

\begin{align}
\varepsilon=& 
\sum_{i=n,p,e,\mu} \frac{m_i^{*4}}{8\pi^2}
\left[
x_i \left(2x_i^2+1\right)\sqrt{x_i^2+1}
-\operatorname{arcsinh}(x_i)
\right] \nonumber\\
&+\frac{1}{2}\left(\nabla \sigma\right)^2
+\frac{1}{2}m_\sigma^2\sigma^2
+\frac{1}{2}\left(\nabla \omega\right)^2
+\frac{1}{2}m_\omega^2\omega^2\nonumber
\\
&+\frac{1}{2}\left(\nabla \rho\right)^2
+\frac{1}{2}m_\rho^2\rho^2
+\frac{1}{2}\left(\nabla \mathscr{A}\right)^2 \nonumber\\
&+c_3\omega^4
+d_3\rho^4
-U(\sigma,\omega,\rho).
\end{align}
\label{eq:energy_density}
Here, the first term corresponds to the local kinetic energy density with $x_i=\nu_i/m_i^*$ (note that for the lepton masses $m_e^*=m_e$ and $m_\mu^*=m_\mu$). 
The pressure is then obtained from the Gibbs-Duhem relation:
\begin{equation}
p=\sum_{i}\mu_i n_i-\varepsilon .
\label{eq:pressure}
\end{equation}
Here, $n_i$ and $\mu_i$ denote the number density and chemical potential of the $i$-th particle species, respectively.
Finally, the NS EoS is obtained by imposing the charge neutrality condition:
\begin{equation}
    \sum_i q_in_i = 0,
\end{equation}
and the $\beta$-equilibrium condition:
\begin{equation}
    \mu_b = \mu_n -q_b\mu_e, \; \mu_\mu = \mu_e.
\end{equation}
Here, we adopt the GM1 parameterization~\cite{Glendenning:1991es} to model NS in our work (see Table.~\ref{tab:GM1}). Detailed description of the EoS and emerging NS properties can be seen in Refs.~\cite{Xia:2022dvw,Niu:2025tvd}.
\begin{table}[ht]
\centering
\caption{GM1 model parameters~\cite{Xia:2022dvw,Niu:2025tvd} used in the present work.}
\label{tab:GM1}
\begin{tabular}{|cc|cc|}
\hline\hline
Mass & Value & Coupling & Value \\
\hline
$m_n$      & 938 MeV           & $g_\sigma$ & 8.78443 \\
$m_p$      & 938 MeV           & $g_\omega$ & 10.60957 \\
$m_\sigma$ & 510 MeV           & $g_\rho$   & 4.09772 \\
$m_\omega$ & 783 MeV           & $g_2$      & $-9.7908$ fm$^{-1}$ \\
$m_\rho$   & 770 MeV           & $g_3$      & $-6.63661$ \\
$m_e$      & 0.511 MeV         & $c_3$      & 0 \\
$m_\mu$    & 105.66 MeV        & $d_3$      & 0 \\
\hline\hline
\end{tabular}
\end{table}

%%%%%%%%%%%%%%%%%%%%%%%%%%%%%%%%%%%%%%%%%%%%%%%%%%%%%%%%%%%%%%%%%%%%%%%%%%%%%%%%%%%%%%%%%
\par 
\begin{figure}[t]
\includegraphics[width=\linewidth]{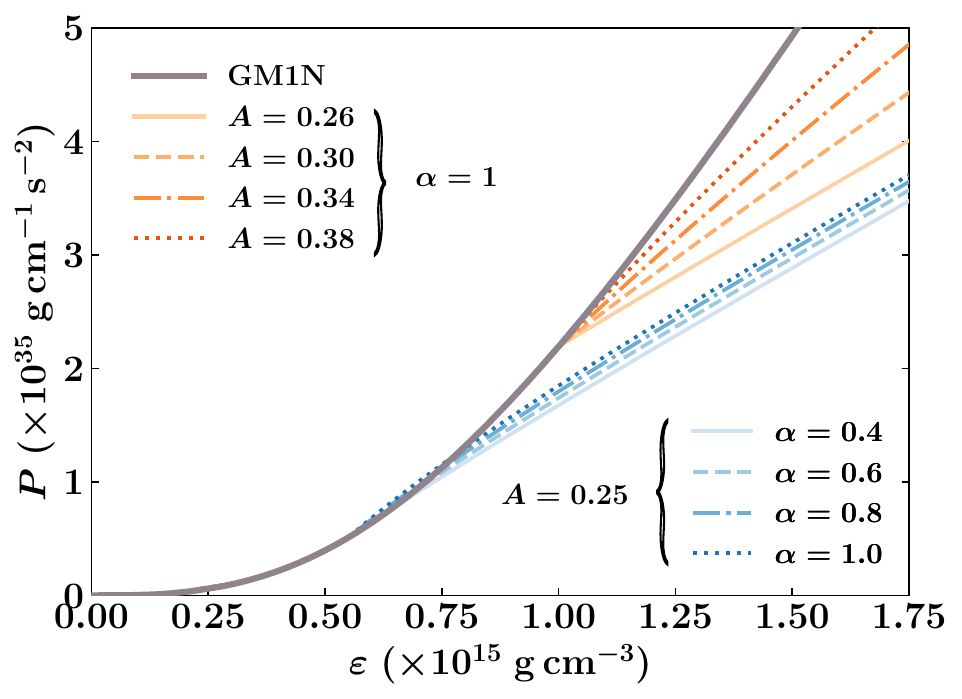}
 \caption{Equation of state of neutron stars (NSs) with DE core. The solid curve represents the pure NS model~\cite{Serot:1979dc}, whereas the dashed curves represent the variation of the Chaplygin fluid (DE) model parameters $A$ and $\alpha$.\label{fig:eos} }
 \end{figure}
Now, we proceed to construct the hybrid EoS consisting of nuclear and DE matter (DENS). The transition between the DE core and the nuclear matter envelope is treated as a sharp interface, such that the pressure and energy density are continuous. The pressure and energy density at the transition point ($r=r_t$) are given as
\begin{align}
p_{\rm core}^{\rm DE}(r_t)&=p_t=p_{\rm env}^{\rm NS}(r_t),
\nonumber\\
\varepsilon_{\rm core}^{\rm DE}(r_t)&=\varepsilon_t =\varepsilon_{\rm core}^{\rm NS}(r_t).
\end{align}
These matching conditions were provided to ensure that there is no surface layer, density discontinuity, or singular stress-energy contribution at the transition region. 

For a fixed values of parameter set ($A,\alpha$), choosing a particular transition pressure and density ($p_t,\varepsilon_t$), allows the remaining DE EoS parameter $B$ to be uniquely determined as
\begin{equation}
B = \left(A\varepsilon_t - p_t\right)\varepsilon_t^{\alpha}.
\end{equation}
Now, varying DE parameters for any transition density generates different 
hybrid star EoSs used for this study. 

\begin{figure}[b]
\includegraphics[width=\linewidth]{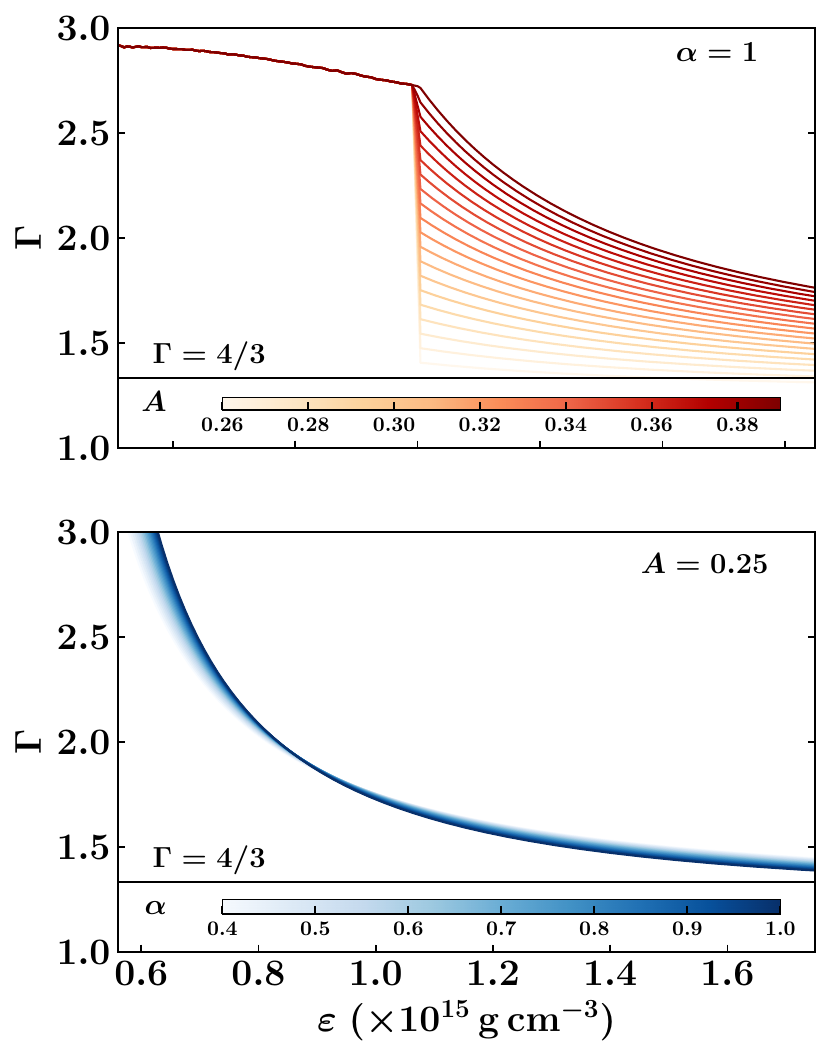}
 \caption{The adiabatic index ($\Gamma$) as a function of the energy density for the EoSs with DE cores.
 \label{fig:gamma}}
 \end{figure}

In Fig.~\ref{fig:eos}, we plot the EoS constructed for models considered by varying the DE parameters $A$ and $\alpha$.
The gray solid curve represents the pure nuclear matter. The coloured curves correspond to hybrid configurations containing a DE core modelled through the Chaplygin gas prescription. The orange and blue curves illustrate the effects of varying $A$ from $0.26$ to $0.38$ and $\alpha$ from $0.4$ to $1.0$, respectively. The two sets of models also differ from each other by the choice of the transition point, where for variation in $A$ ($\alpha=1$) we chose $\varepsilon_t = 1.0\times10^{15}$ g cm$^{-3}$ and for variation in $\alpha$ ($A=0.25$) the choice is $\varepsilon_t=0.55\times10^{15}$ g cm$^{-3}$. In the former case the DE fluid is distributed in the inner core of the NS, while in the latter case, it extends throughout most of the stellar core. 
For both the scenarios, we find that reducing the values of DE parameter $A$ or $\alpha$ results in softening of the EoSs. 
For all the models taken, it is verified that they obey the causality condition (square of speed of sound $c_s^2<1$). 
Additionally, we also examine the stability of the EoSs considered by analysing the condition for dynamic stability in terms of adiabatic index 
\begin{equation}
    \Gamma =\frac{\varepsilon+P}{P} \left( \frac{dP}{d\varepsilon} \right) >4/3,
\end{equation}
as shown in Fig.~\ref{fig:gamma}.
We find that all the models considered satisfy the above mentioned relation throughout the density range. 
Although the onset of DE gives rise to a decrease in $\Gamma$ near $\varepsilon_t$, no violation of the dynamic stability criterion is observed. For our analysis, the values of the Chaplygin parameters $A$ and $\alpha$ are chosen such that the resulting EoSs satisfy the stability criterion. 
%%%%%%%%%%%%%%%%%%%%%%%%%%%%%%%%%%%%%%%%%%%%%%%%%%%%%%%%%%%%%%%%%%%%%%%%%%%
\section{Formalism}
\label{sec:GR}
Now, we present the general relativistic prescriptions employed to obtain the stellar structural and non-radial oscillation properties. 
%%%%%%%%%%%%%%%%%%%%%%%%%%%%%%%%%%%%%%%%%%%%%%%%%%%%%%%%%%%%%%%%%%%%%%%%%%%
\subsection{Stellar structure}
The line element for a static spherically symmetric stellar equilibria is given as
\begin{align}
   ds^{2}&=e^{\phi(r)} dt^{2}-e^{ \lambda(r)} dr^{2}-r^{2}\left(d\theta ^{2}+\sin ^{2}\theta d\varphi ^{2}\right);
\end{align}
where, $\phi(r)$ and $\lambda(r)$ are the metric functions. The solutions obtained by solving the Einstein field equations for the given metric are the Tolman–Oppenheimer–Volkoff (TOV) equations \cite{Tolman1939,Oppen1939}:\begin{align}
     \label{P-tov}
    \frac{dp}{dr}&=-\frac{\left(\varepsilon+p\right )
        \left(m+4\pi r^3 p\right)}{r(r-2m)},\\
        \frac{dm}{dr}&= 4\pi r^2 \varepsilon;
        \label{M-tov}
\end{align}
%\par
describing the stellar structure. The TOV equations are integrated outward from the stellar center to the surface to determine the gravitational mass $M$ and radius $R$ of the stellar equilibrium, with the central boundary conditions
\begin{equation}
p(0)=p_c=p(\varepsilon_c),\qquad m(0)=0,
\end{equation}
where $\varepsilon_c$ being the central density. The stellar surface $R$ is located where the pressure vanishes, $p(R)=0$. 
The metric function $\phi$ is obtained by integrating the equation
\begin{align}
    \frac{d\phi}{dr} &= -\frac{1}{(\varepsilon+p)}\,\frac{dp}{dr}
\end{align}
simultaneously with the TOV equations from the center to the surface of the star, using the boundary condition $\phi(0)=0$. By varying the central density $\varepsilon_c$ and repeating the integration, one obtains a sequence of equilibrium models with corresponding gravitational mass $M$ and radius $R$. 

\subsection{Linear perturbation formalism for even parity perturbations}
We analyse the non-radial oscillations of NSs with a DE core using the linearised perturbation formalism~\cite{Lindblom:1983ps,Detweiler:1985zz}. The perturbations are categorized into \textit{even-parity} (polar) and \textit{odd-parity} (axial) modes based on their transformation properties under parity transformations. In this study, we focus on the even-parity perturbations, which are associated with fluid oscillations and GW emission. 
For a fully relativistic analysis, we need to consider both the perturbations in the space-time metric and the fluid variables. The perturbation functions are expanded in terms of spherical harmonics with indices ($\ell, m$), and we adopt the \textit{Regge-Wheeler gauge}~\cite{Regge:1957td} to simplify the analysis. Now, the metric perturbations are expressed as
\begin{align}
h_{\mu\nu} &= - r^{\ell} \mathbb{H}_{\mu\nu}
Y_{\ell}^{m} \, e^{i\omega t} \,; \\
\newline \nonumber\\
\textrm{with,}\qquad \mathbb{H}_{\mu\nu} &=
\begin{pmatrix}
e^{\phi} H_{0} & i \omega r H_{1}& 0 & 0 \\
i \omega r H_{1} & e^{\lambda} H_{2} & 0 & 0 \\
0 & 0 & r^{2} K & 0 \\
0 & 0 & 0 & r^{2}\sin^{2}\theta \, K
\end{pmatrix}\cdot
\nonumber
\end{align}
Here, the metric perturbations $H_0,\,H_1,\,H_2$ and $K$ are functions of $r$.
 The components of the displacement vector ($\xi^\mu$) are expressed in terms of the fluid perturbation functions $V$ and $W$, as
\begin{subequations}
\begin{align}
\xi^{r} &= r^{\ell-1} e^{-\lambda/2} W Y_{\ell}^{m} e^{i\omega t}, \\
\xi^{\theta} &= -r^{\ell-2} V \,\partial_{\theta} Y_{\ell}^{m} e^{i\omega t}, \\
\xi^{\phi} &= -\frac{r^{\ell-2}}{\sin^{2}\theta}\,
V \,\partial_{\phi} Y_{\ell}^{m} e^{i\omega t}\cdot
\end{align}
\end{subequations}
The homogeneous linear equations for the perturbation functions obtained by solving field equations are thus given by~\cite{Lindblom:1983ps,Detweiler:1985zz}:
\begin{subequations}\label{eq:diff-fgR}
\begin{align}
\frac{dH_1}{dr}
&=
-r^{-1}
\left[
\ell+1+\frac{2Me^{\lambda}}{r}
+4\pi r^2 e^{\lambda}(p-\varepsilon)
\right]H_1 \nonumber\\
&\qquad +e^{\lambda}r^{-1}
\Bigl[
H_0+K-16\pi(\varepsilon+p)V
\Bigr],
\\
\frac{dK}{dr}
&=
r^{-1}H_0
+\frac{\ell(\ell+1)}{2r}H_1
-\left[
\frac{\ell+1}{r}
-\frac{\phi'}{2}
\right]K\nonumber\\
&\qquad -8\pi(\varepsilon+p)e^{\lambda/2}r^{-1}W,
\\
\frac{dW}{dr}
&=
-(\ell+1)r^{-1}W
+re^{\lambda/2}
\Big[
e^{-\phi/2}\Gamma^{-1}p^{-1}X
\nonumber\\
&\qquad -\ell(\ell+1)r^{-2}V
+\frac{1}{2}H_0
+K
\Big],
\\
\frac{dX}{dr}
&=
-\ell r^{-1}X
+\frac{(\varepsilon+p)e^{\phi/2}}{2}
\Biggl[
\left(
\frac{1}{r}
+\frac{\phi'}{2}
\right)H_0
\nonumber\\
&\qquad
+\left(
r\omega^2e^{-\phi}
+\frac{\ell(\ell+1)}{2r}
\right)H_1
+\left(
\frac{3}{2}\phi'
-\frac{1}{r}
\right)K\nonumber\\
&\qquad-\ell(\ell+1)r^{-2}\phi'V
-2r^{-1}
\Big(
4\pi(\varepsilon+p)e^{\lambda/2}
\nonumber\\
&\qquad +\omega^2e^{\lambda/2-\phi}
-\frac{r^2}{2}
\bigl(
e^{-\lambda/2}r^{-2}\phi'
\bigr)'
\Big)W
\Biggr].
\end{align}
\end{subequations}
Here, (') denotes differentiation with respect to $r$ and $\Gamma$ is the adiabatic index. 
Further, the perturbation equations for $H_0$ and $X$ are expressed as
\begin{align}
 H_0
&= \left[
\frac{r e^{-\lambda}}{2} (r \phi' - 2) + (n+1) r
\right]^{-1}\Bigg\{ 8\pi r^3 e^{-\phi/2} X
\nonumber \\
&\quad  + r^2 e^{-\lambda} \left(
\omega^2 r e^{-\phi} - \frac{n+1}{2} \phi'
\right)H_1 \nonumber \\
&\quad + \left[
nr - \omega^2 r^3 e^{-\phi}
- \frac{1}{4} r^2 e^{-\lambda} \phi' (r \phi' - 2)
\right] K\Bigg\}, \\
X &= \omega^2 (\varepsilon + p)\, e^{-\phi/2} V
- \frac{1}{r} p' e^{(\phi - \lambda)/2} W \nonumber \\
&\quad + \frac{1}{2} (\varepsilon + p)\, e^{\phi/2} H_0 .
\end{align}
Here, $n$ is the angular parameter given by $n = (l-1)(l+2)/2$.
To solve the above given system of ordinary differential equations, we need to impose appropriate boundary conditions. At the center of the star ($r=0$), regularity conditions require that the perturbation functions behave as~\cite{Lindblom:1983ps,Detweiler:1985zz}
\begin{align}
X(0) &=
\left\{
\left[
\frac{4\pi}{3}(\varepsilon_0 + 3p_0)
- \frac{\omega^2}{l\,} e^{-\phi_0}
\right] W(0)
+ \frac{K(0)}{2}
\right\} \nonumber\\
& \quad \times  (\varepsilon_0 + p_0)\, e^{\phi_0/2}, \\[6pt]
H_1(0) &= \left\{lK(0) + 8\pi(\varepsilon_0 + p_0)W(0)\right\}/(n + 1), \\[6pt]
H_0(0)& = K(0).
\end{align}
At the stellar surface ($r=R$), the Lagrangian perturbation of the pressure must vanish, which leads to the condition
\begin{equation}
X(R) = 0.      
\end{equation}
Finally, the perturbation functions must match continuously to the Zerilli function in the exterior vacuum region, which satisfies the Zerilli equation~\cite{Zerilli:1970se}. This matching condition ensures that the perturbations are physically consistent across the stellar surface and allows us to determine the eigen frequencies of the oscillation modes.
The \textit{Zerilli equation} governing the exterior perturbations is given by~\cite{Zerilli:1970se}
\begin{equation}
\frac{d^2 Z(r^*)}{d r^{*2}} + \left[\omega^2 - V_Z(r^*)\right] Z (r^*) = 0,  \label{eq:Zerili}
\end{equation}
with the tortoise coordinate ($r^*$) defined as
\begin{equation}
r^* = r + 2M \log\!\left(\frac{r}{2M} - 1\right). 
\end{equation}
The effective potential $V_Z(r^*)$ in the Zerilli equation is given by 
\begin{align}
V_Z(r^*) =&
\frac{2(1 - 2M/r)}{r^3 (n r + 3M)^2}
\Big[
n^2 (n+1) r^3
+ 3 n^2 M r^2\nonumber\\
&
+ 9 n M^2 r
+ 9 M^3
\Big].
\end{align}
The Zerilli function $Z(r^*)$ and its derivative can be expressed in terms of the metric perturbations $H_0(r)$ and $K(r)$ by the linear transformation~\cite{fackerell1971solutions} 
\begin{equation}
\begin{pmatrix}
0 & 1 \\
a & b
\end{pmatrix}
\begin{pmatrix}
H_0 \\
K
\end{pmatrix}
=
\begin{pmatrix}
g & 1 \\
h & k
\end{pmatrix}
\begin{pmatrix}
Z(r^*) \\
dZ/dr^*
\end{pmatrix}. \label{eq:trans}
\end{equation}
The transformation coefficients $(a,\,b,\,g,\,h,\,k)$ are defined as follows:
\begin{subequations}
\begin{align}
a(r) &= -\frac{n r + 3M}
{\omega^2 r^2 - (n+1)M/r}, \\[6pt]
b(r) &= \frac{n r - \omega^2 r^4 + M(r - 3M)}
{(r - 2M)\left[\omega^2 r^2 - (n+1)M/r\right]}, \\[6pt]
g(r) &= \frac{n(n+1)r^2 + 3n M r + 6M^2}
{r^2 (n r + 3M)}, \\[6pt]
h(r) &= \frac{-n r^2 + 3n M r + 3M^2}
{(r - 2M)(n r + 3M)}, \\[6pt]
k(r) &= -\frac{r^2}{r - 2M}\cdot
\end{align}
\end{subequations}

The asymptotic solutions to the Zerilli equation consist of two linearly independent modes: an outgoing wave $Z_{-}(r^*)$ and an incoming wave $Z_{+}(r^*)$, represented by corresponding power series
\begin{subequations}\label{eq:Z+-}
\begin{align}
Z_{-}(r^*) &= e^{-i\omega r^*} \sum_{j=0}^{\infty} \alpha_j \, r^{-j}, \\
Z_{+}(r^*) &= e^{i\omega r^*} \sum_{j=0}^{\infty} \bar{\alpha}_j \, r^{-j};
\end{align}
\end{subequations}
where, the coefficient $\bar{\alpha}_j$ is the complex conjugate of $\alpha_j$. Substituting these expansions into the Zerilli equation and keeping the terms up to $\mathcal{O}(r^{-2})$ yields
\begin{align}
\alpha_1 &= -i\omega^{-1}(n + 1)\alpha_0, \\
\alpha_2 &= -\frac{1}{2\omega^2}
\left[
n(n + 1) - 3iM\omega\left(1 + \frac{2}{n}\right)
\right]\alpha_0;
\end{align}
with, $\alpha_0$ being an arbitrary constant. The general Zerilli solution combines both the modes as
\begin{equation}
Z(r^*) = A_{\mathrm{out}}\, Z_{-}(r^*) 
        + A_{\mathrm{in}}\, Z_{+}(r^*);
\end{equation}
with, $A_{\mathrm{out}}$ and $A_{\mathrm{in}}$ representing the amplitudes of the outgoing and incoming components, respectively.
In the exterior region, the transformation equations, \textit{i.e.}; Eq.~\eqref{eq:trans}, are solved using the values of $H_0$ and $K$ obtained using their surface boundary conditions. The resulting exterior solution is then matched with the asymptotic limit to determine the coefficients $A_{\mathrm{out}}$ and $A_{\mathrm{in}}$. 

The Zerilli equation is integrated outward from the surface  ($r=R$) to a large distance ($r \approx 50\,Re\,\omega^{-1}$), where the asymptotic behavior of the solution is analysed.
Here, the surface boundary condition requires that the solution be purely outgoing, which implies $A_{\mathrm{in}} = 0$. This condition leads to a discrete set of complex eigen frequencies $\omega = \omega_r + i\,(1/\tau)$ that correspond to the quasi-normal modes of the star. The real part of $\omega$ gives the oscillation frequency, while the imaginary part determines the damping time of the mode due to GW emission.

Thus, we can calculate the even parity modes of any order by solving the above described eigen value prescription. The $f$-modes are non-radial modes with $n=0$ nodes, while modes with $n=1,2,3,..$ are referred to as pressure $p$-modes. In this work, we study the $f$-modes in detail and also briefly explore  the higher overtones ($p$-modes). 
%%%%%%%%%%%%%%%%%%%%%%%%%%%%%%%%%%%%%%%%%%%%%%%%%%%%%%%%%%%%%%%%%%%%%%%%%%%%%%%%%%%%%%%%%%%%%%%%%%%%%%%%%%%%%%%
\subsection{Tidal deformability}
The detection of GWs from binary NS mergers has opened a unique window in probing the internal structure of NSs. The strong gravitational field created in the late stage of the inspiral phase of a binary merger results in quadrupole deformation of the stellar structure. This can be quantified through the dimensionless tidal deformability $\Lambda$~\cite{Hinderer2008,Hind2010}. 

In a static external quadrupolar tidal field $\mathcal{E}_{ij}$, the tidal deformability of a static, spherically symmetric star to a linear order is defined as $Q_{ij} = -\Lambda\mathcal{E}_{ij},$ where $Q_{ij}$ is the star's induced quadrupole moment. The $l=2$ tidal Love number $k_2$ can be written in terms of $\Lambda$ and $R$ as $k_2 = (3/2)\Lambda R^{-5}$. To calculate $k_2$, we solve the following set of coupled first order differential equations \cite{Hinderer2008,Hind2010}:
\begin{eqnarray}
\frac{dH}{dr}&=& \beta,\\
\frac{d\beta}{dr}&=&2 \left(1 - 2\frac{m}{r}\right)^{-1} H\left\{-2\pi
  \left[5\varepsilon+9
    p+\frac{d\varepsilon}{dp}(\varepsilon+p)\right]\phantom{\frac{3}{r^2}} \right. \nonumber\\
&&\, + \left. \frac{3}{r^2}+2\left(1 - 2\frac{m}{r}\right)^{-1}
  \left(\frac{m}{r^2}+4\pi r p\right)^2\right\}\\
&&\,+\frac{2\beta}{r}\left(1 -
  2\frac{m}{r}\right)^{-1}\left\{-1+\frac{m}{r}+2\pi r^2
  (\varepsilon-p)\right\}.\nonumber
\end{eqnarray}
Here, $H(r)$ is the metric perturbation function that describes the stellar response to the external tidal field and $\beta(r)=dH/dr$.    
The above set of coupled equations are solved simultaneously with the TOV equations for pressure and mass (Eqs.~\ref{P-tov} and \ref{M-tov}), by setting $H(r) = a_0r^2$ and integrating from the center to the surface of the star.
The constant $a_0$ is arbitrary and cancels out in the expression for the Love number. 
Now, for $l=2$, the Love number ($k_2$) in terms of $y = { R \,\beta(R)}/{H(R)}$ and compactness $C=M/R$ is obtained as~\cite{Hind2010}:
\begin{eqnarray}
k_2 &=& \frac{8C^5}{5}(1-2C)^2[2+2C(y-1)-y]\\
&& \quad \times \bigg\{2C[6-3y+3C(5y-8)]\nonumber\\
      && \quad +\,4C^3[13-11y+C(3y-2)+2C^2(1+y)]\nonumber\\
      && \quad +\,3(1-2C)^2[2-y+2C(y-1)] \ln(1-2C)\bigg\}^{-1}.\nonumber
\label{eq:k2}
\end{eqnarray}

%%%%%%%%%%%%%%%%%%%%%%%%%%%%%%%%%%%%%%%%%%%%%%%%%%%%%%%%%%%%%%%%%%%%%%%%%%%
\subsection{Gravitational wave energy and strain}
The excitation of $f$-modes leads to the emission of GWs. The ability of these modes to strongly couple with gravitational radiation makes the calculation of the energy stored in the stellar oscillations and the corresponding GW strain amplitude relevant. 
In a weak field limit, the energy contained in a stellar oscillation mode is computed as~\cite{Thorne1969III} 
\begin{align}\label{eq:E_osc}
E_{\rm osc}
=&
\frac{1}{2}\omega_r^{2}
\int_{0}^{R}
(\varepsilon+p)e^{(\lambda-\phi)/2}
r^{4}\Big\{
\bigl[W_{\omega}(r)\bigr]^{2} \nonumber\\
&+l(l+1)\bigl[V_{\omega}(r)\bigr]^{2}
\Big\}\,
dr .
\end{align}
Here, $W_{\omega}$ and $V_{\omega}$ are the fluid perturbation functions that correspond to the real values of $\omega$. 

Another quantity of direct observational relevance is the GW strain amplitude $h_{+}$ related to the signal strength measured by the detector.  We estimate the GW strain amplitude from the relation~\cite{Zheng:2025xlr}
\begin{equation}
h_{+}
=
\frac{\sqrt{30\,E_{\rm GW}/\tau_{\rm GW}}}{\omega_r D};
\end{equation}
where, $\omega_r$ and $\tau_{\rm GW}$ denote the $f$-mode frequency and damping time, respectively. Also, $E_{\rm GW}$ is the energy emitted in GWs and $D$ is the distance between the source and the detector.
This estimate can give us an idea about GW strain that could be possibly measured using detectors like LIGO and Virgo. 
%%%%%%%%%%%%%%%%%%%%%%%%%%%%%%%%%%%%%%%%%%%%%%%%%%%%%%%%%%%%%%%%%%%%%%%%%%%%%%%%%%%%%%%%%%%%%%%%%%%%%%%%%%%%%%%%%%%%%%%%

\section{Results and discussion}
\label{sec:results}

We present the static stellar structure and the non-radial fundamental oscillation properties of NSs with a DE core (DENS) described by the EoSs obtained in Section~\ref{sec:eos}. We also compute the tidal deformability and the GW strain for these configurations. 
We obtain these results by numerically solving the corresponding stellar structure and oscillation equations (Section~\ref{sec:GR}) using the hybrid EoSs. We vary the DE parameters $A$ ($0.24-0.38$) and $\alpha$ ($0.4-1.0$) and the transition energy density $\varepsilon_t$ (resulting in different DE core radii) in this study. 

\begin{figure}[t]
\includegraphics[width=\linewidth]{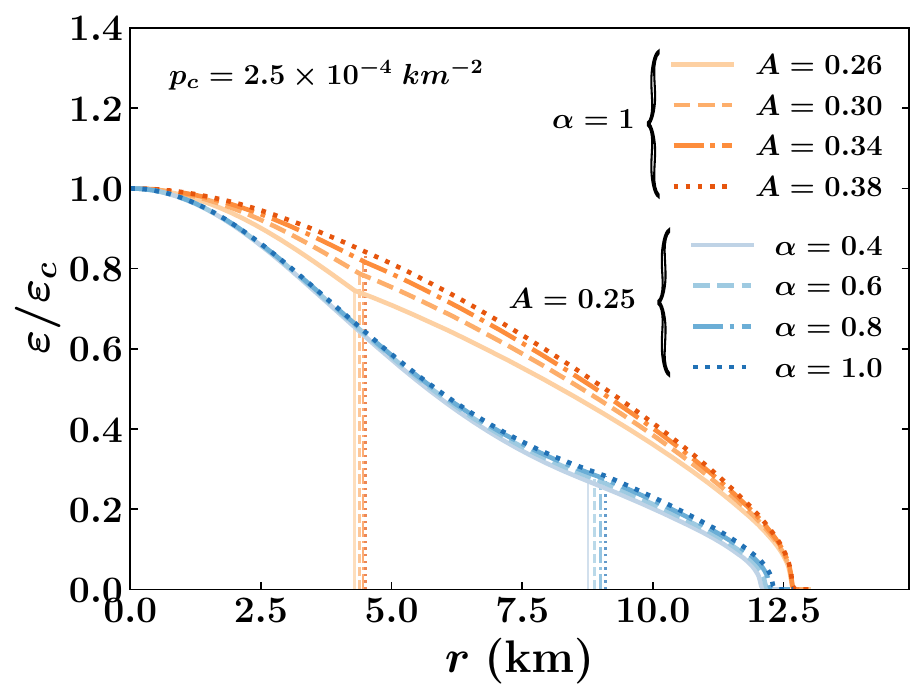}
 \caption{Normalised stellar energy density profiles (with a fixed center pressure $p_c = 2.5\times 10^{-4}$ km$^{-2}$)
 as a function of radius $r$ for varied values of DE EoS parameters. The vertical lines denote the DE core radii ($r_t$). \label{fig:rho-r}}
 \end{figure}

%%%%%%%%%%%%%%%%%%%%%%%%%%%%%%%%%%%%%%%%%%%%%%%%%%%%%%%%%%%%%%%%%%%%%%%%%%%%%%%%%%%%
First, we numerically solve the TOV Eqs.~\eqref{P-tov} and \eqref{M-tov} from the center to the surface of the star.
In Fig.~\ref{fig:rho-r}, we plot the normalised energy density profiles as a function of radius $r$ for a fixed central pressure $p_c = 2.5\times 10^{-4}$ km$^{-2}$, by varying the EoS parameters. From the center to the surface of the star, the density profiles decrease monotonically. The density profiles are found to decrease with increasing $A$, with the maximum reduction occurring near the respective DE core radii, although the reduction becomes minimal toward the stellar surface. 
Whereas, we find that the density profiles obtained by varying $\alpha$ remain close to one another. The vertical lines depict DE core radii obtained for each profile. 
We observe that for a higher (lower) value of transition density $\varepsilon_t=1.0\times10^{15}$ g cm$^{-3}$ ($\varepsilon_t=0.55\times10^{15}$ g cm$^{-3}$), the resulting DE core radius has lower (higher) value. The difference between the DE core radii in the two cases is approximately $\sim4$ km. 
Further, the DE core radius ($r_t$) is found to increase slightly for larger $A$ and $\alpha$ values.

%%%%%%%%%%%%%%%%%%%%%%%%%%%%%%%%%%%%%%%%%%%%%%%%%%%%%%%%%%%%%%%%%%%%%%%%%%%%%%%%%555
 
\begin{figure}[t]
\includegraphics[width=\linewidth]{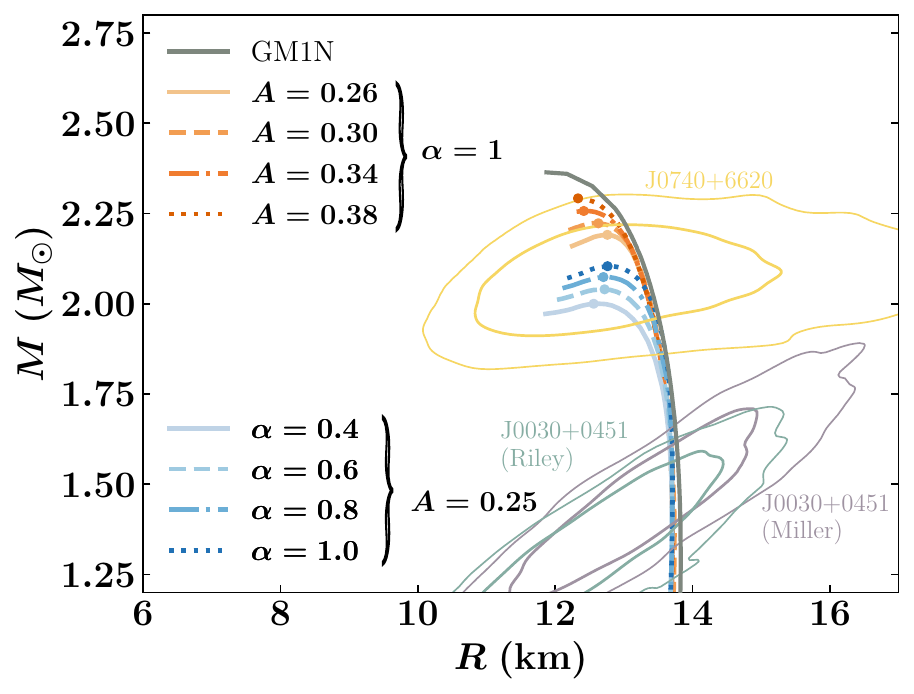}
\caption{Mass-radius relations for the models shown in Fig.~\ref{fig:eos}.  The constraints on $M-R$ plane prescribed from the NICER experiment for PSR J0030+0451~\cite{Riley:2019yda, Miller:2019cac} and PSR J0740+6620 \cite{Fonseca:2021wxt}. 
\label{fig:MR}}
\end{figure}

The mass-radius relations obtained for the models considered are plotted in Fig.~\ref{fig:MR}. We also show the stellar profile of pure nuclear matter (GM1N) for comparison. The filled circles correspond to the maximum mass profiles. We find that an increase in the EoS parameter $A$ from $0.26$ to $0.38$ leads to higher maximum masses and smaller corresponding radii. 
As the choice of a higher value of the transition density ($\varepsilon_t=1.0\times10^{15}$ g cm$^{-3}$) confines the DE matter to the innermost region of the stellar profile, the mass and radius values show considerable change only in the higher mass region. 
Next, as we increase the value of $\alpha$ from $0.4$ to $1.0$ (with $\varepsilon_t=0.55\times10^{15}$ g cm$^{-3}$), the maximum mass and the corresponding radius are seen to increase. 
We find that the presence of DE over larger regions of the star brings visible changes in the mass-radius profile from $M\sim1.70M_\odot$ itself. Thus, we see that inclusion of DE in the core of a NS decreases the maximum mass of the star compared to a pure NS. The $M-R$ curves indicate that the compactness ($C=M/R$) increases with stellar mass, as the radius remains nearly constant or slightly decreases over the given mass range. 
We also compare our results with the mass-radius constraints given by NICER for PSR J0030+0451~\cite{Riley:2019yda, Miller:2019cac} and PSR J0740+6620~\cite{Fonseca:2021wxt}. We find that the mass-radius values of the models considered in our work are in close agreement with these observational constraints. 
 
\begin{figure}[t]
\includegraphics[width=\linewidth]{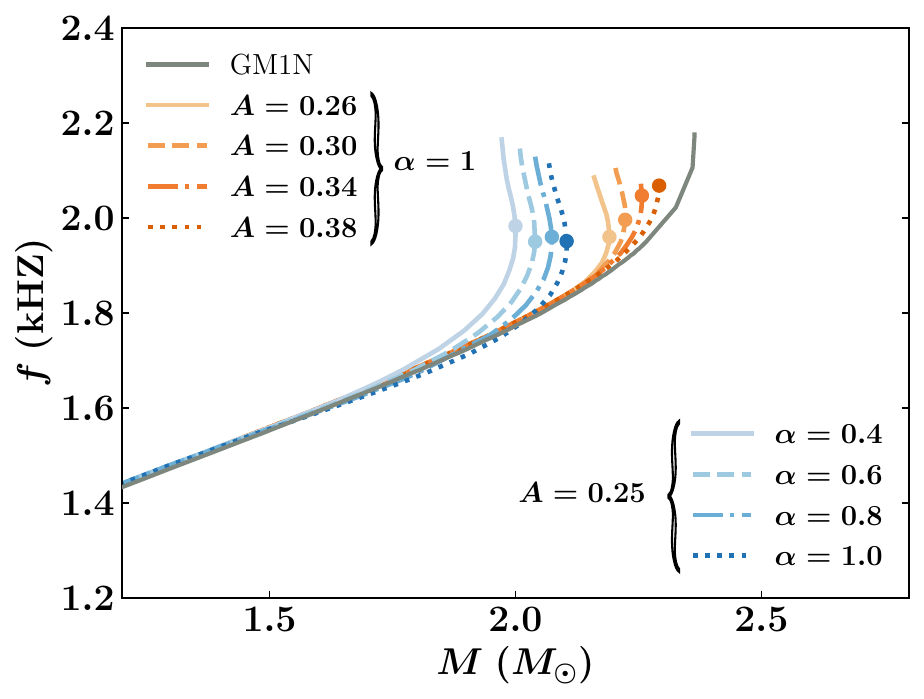}
\caption{$f$-mode frequencies as a function of stellar mass $M$ for the models considered.\label{fig:fM}}
\end{figure}

%%%%%%%%%%%%%%%%%%%%%%%%%%%%%%%%%%%%%%%%%%%%%%%%%%%%%%%%%%%%%%%%%%%%%%%%%%%%%%%%%%%
We now proceed to study the non-radial $f$-mode oscillations of NSs containing a DE core. In Fig.~\ref{fig:fM}, we plot the $f$-mode frequency as a function of stellar mass by varying the EoS parameters $A$ and $\alpha$. 
For the models obtained by varying $A$, the frequencies deviate from pure NS model only in the high mass region. This trend is consistent with the corresponding mass-radius relation, where the effects of DE core on the stellar structure become noticeable only for massive stellar configurations. In contrast, varying the parameter $\alpha$ results in a $f$-mode spectrum that extends over a broader range of stellar mass. 
The frequency corresponding to the maximum mass ($f_{max}$) decreases slightly with a reduction in $A$ and lie in the range of (1.8, 2.1) kHz. On the other hand, we see that $f_{max}$ remains almost equal for any value of $\alpha$ considered. 
Further, it is interesting to note that the models obtained by varying $\alpha$ span a narrow range of frequencies ($\leq1.9$kHz), whereas those obtained by varying $A$ extend to slightly higher values ($\leq 2.1$ kHz).

\begin{figure}[t]
\includegraphics[width=\linewidth]{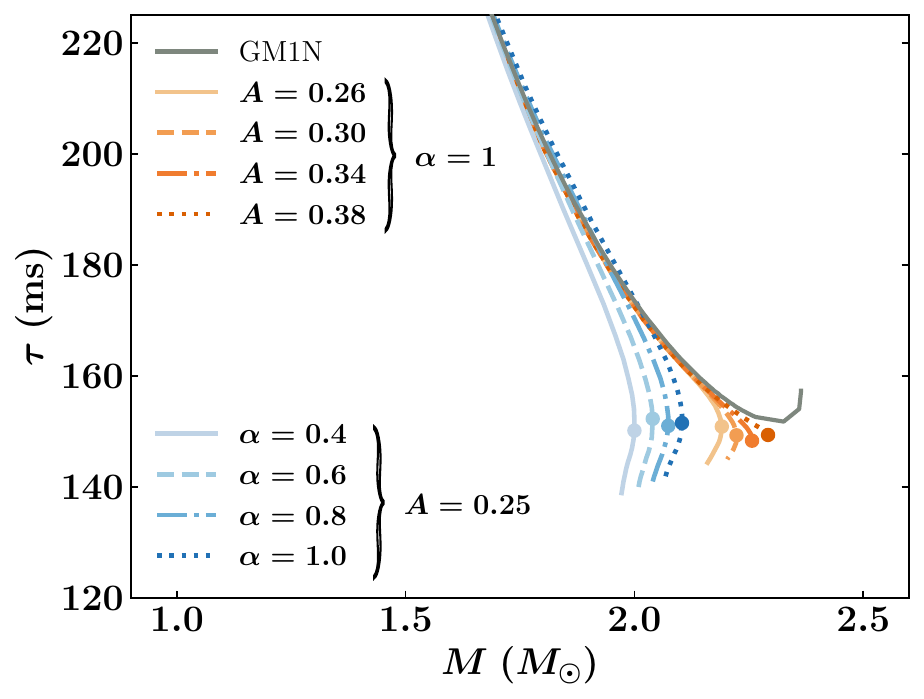}
\caption{Variation of damping time ($\tau$) of $f$-mode oscillations with stellar mass $M$. \label{fig:tM}}
\end{figure}

%%%%%%%%%%%%%%%%%%%%%%%%%%%%%%%%%%%%%%%%%%%%%%%%%%%%%%%%%%%%%%%%%%%%%%%%%%%%%%%%%%%
Next, we examine the damping time $\tau$ of the $f$-mode as a function of $M$ for the models considered and are plotted in Fig.~\ref{fig:tM}. We find a monotonic decrease of $\tau$ with increasing stellar mass for all the models shown. 
The damping time decreases from approximately $400$ ms for low-mass stars ($M\sim1.2\,M_\odot$) to about $140$-$150$ ms near the maximum-mass configurations.
This indicates that more massive configurations damp the $f$-mode oscillations on increasingly shorter timescales. For the models obtained by varying $A$, the $\tau$ profiles remain identical almost throughout the stellar masses and show deviation only near the maximum mass region. This behavior is inline with the corresponding mass-radius profiles, where the influence of DE becomes significant only in the high-mass region. The corresponding damping times near the maximum mass configurations lie in the narrow interval of $\sim 145$-$155$ ms. Meanwhile, the models obtained by varying $\alpha$ exhibit systematically larger $\tau$ values over a broad range of stellar masses. This demonstrates that the effect of the DE component is not confined to the high-mass regime but extend throughout the stellar profiles ($M>1.7M_\odot$). Around $M\simeq2\,M_\odot$, the damping times span a considerably wider interval $\sim 150$-$180$ ms, depending on the value of $\alpha$.

\begin{figure}[t]
\includegraphics[width=\linewidth]{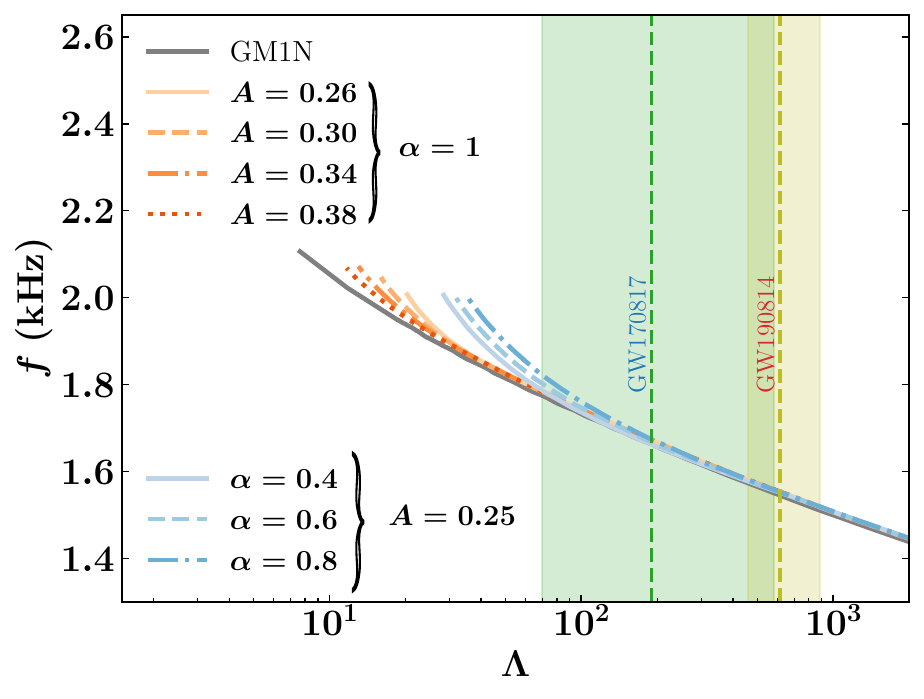}
\caption{$f$-mode frequencies as a function of dimensionless tidal deformability $\Lambda$. \label{fig:fL}}
\end{figure}

%%%%%%%%%%%%%%%%%%%%%%%%%%%%%%%%%%%%%%%%%%%%%%%%%%%%%%%%%%%%%%%%%%%%%%%%%%%%%%%%%%%
 In Fig.~\ref{fig:fL}, we depict the variation of $f$-mode frequencies with the tidal deformability $\Lambda$. The $f-\Lambda$ values are inversely related and the frequency decreases from $\sim2.1$ to $1.45$ kHz as $\Lambda$ increases from $\sim10$ to $10^3$. For the increment in $A$, the frequencies are shifted to higher values particularly in the low-$\Lambda$ regime ($\Lambda\lesssim50$).
In contrast, we find that variation in $\alpha$ results in higher frequencies from $\Lambda\sim 200$ itself.
Further, we observe that for a fixed value of $\Lambda$, the frequency increases (decreases) with increasing A ($\alpha$). We also show the observational measurements from GW170817 ($70\leq\Lambda_{1.4}\leq580$)~\cite{LIGOScientific:2018cki} and GW190814  ($458\leq\Lambda_{1.4}\leq 889$)~\cite{LIGOScientific:2020zkf}. These constraints correspond to characteristic frequency ranges of approximately $1.55-1.80$ kHz and $1.50-1.60$ kHz, respectively. 

\begin{figure}[b]
\includegraphics[width=\linewidth]{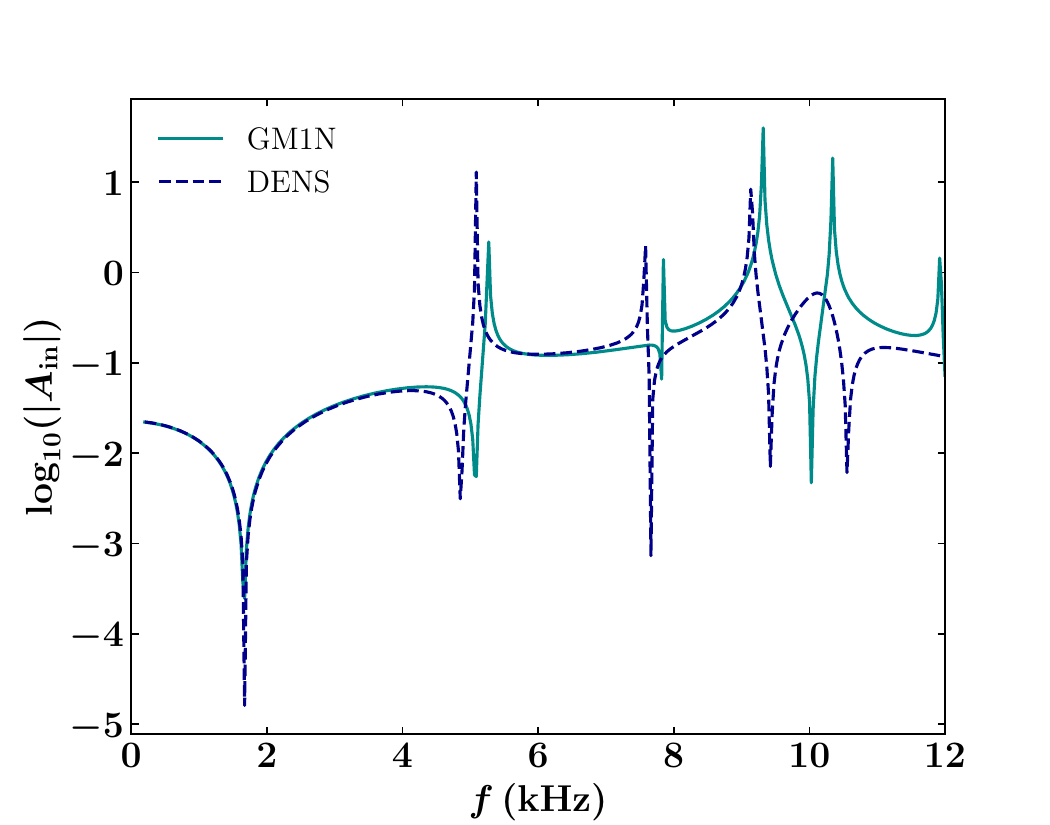}
\caption{The amplitude of incoming GWs ($|A_{\rm in}|$) as a function of real oscillation frequency for pure NS (GM1N) and NS with a DE core (DENS). \label{fig:Ain}}
\end{figure}

%%%%%%%%%%%%%%%%%%%%%%%%%%%%%%%%%%%%%%%%%%%%%%%%%%%%%%%%%%%%%%%%%%%%%%%%%%%%%%%%%%%
Next, we plot the incoming wave amplitude $|A_{\rm in}|$ as a function of the purely real oscillation frequency $\omega_r$ ($f=2\pi\omega_r$) in Fig.~\ref{fig:Ain}. The quasi-normal mode (QNM) frequencies are identified using the shooting method. For a given trial value of the complex eigen frequency, we numerically solve the perturbation equations (Eq.~\ref{eq:diff-fgR}) from the stellar center to the surface by imposing regularity conditions on the fluid and metric perturbations. Near the surface of the star, the interior solutions are matched continuously to the Zerilli function in the exterior Schwarzschild spacetime. We then integrate the Zerilli equation from the stellar surface ($r=R)$ to a large distance ($r=50R$). Here, the asymptotic solution takes the form of Eq.~\ref{eq:Zerili}, which is expressed in terms of outgoing ($A_{\rm out}$) and incoming ($A_{\rm in}$) GW amplitudes. We aim to identify the frequency of QNMs, which consists of purely outgoing radiation, {\it i.e.,} $A_{\rm in}=0$. This is achieved numerically  by varying the trial frequency over a range of values until the quantity $|A_{\rm in}|$ attains its minimum value. The corresponding frequency is the eigen mode frequency. In Fig.~\ref{fig:Ain}, we demonstrate this numerical procedure by showing the variation of $|A_{\rm in}|$ with real frequency values.  
At discrete frequencies we observe distinct minima for the configurations. 
We find that the oscillation spectrum is strongly dependent on the interior composition of the stellar model considered. 
From the figure, we observe a comparatively sharp minima for hybrid star (DENS), indicating a more pronounced identification of the eigen frequencies. 
The estimated frequency shows only a marginal difference with that of pure NS (GM1N) in the low frequency region. 
Using these minima values, now, we can numerically obtain the complex eigen frequencies for all the possible non-radial modes present.  
In Table.~\ref{tab:DENS}, we show the identified complex eigen frequencies for the fundamental $f$- and subsequent $p$-modes of DENS in the frequency range of $f\leq 12$ kHz. We also give the frequencies estimated for GM1N in Table.~\ref{tab:gm1n} for comparison. 

\begin{table}[t]
\centering
\caption{Complex quasi-normal mode frequencies for the neutron star with a DE core (DENS) for $\ell=2$ and central pressure $p_c=8.15\times10^{-5}\,\mathrm{km}^{-2}$.
The corresponding stellar model has radius $R=13.095\,\mathrm{km}$ and mass $M=1.48\,M_{\odot}$.}
\label{tab:DENS}
\small
\begin{tabular}{lccccc}
\hline\hline
Mode & Re($f$) & Im($f$) & $\tau$ & $\omega_rM$ & $\omega_iM$ \\
     & (kHz) & (kHz) & (ms) & & \\
\hline
$f$   & 1.674 & $6.590\times10^{-4}$ & 241.5 & 0.077 & $3.018\times10^{-5}$ \\
$p_1$ & 4.842 & $4.781\times10^{-5}$ & $3.329\times10^{3}$ & 0.222 & $2.189\times10^{-6}$ \\
$p_2$ & 7.386 & $9.507\times10^{-7}$ & $1.674\times10^{5}$ & 0.338 & $4.354\times10^{-8}$ \\
$p_3$ & 9.117 & $2.334\times10^{-7}$ & $6.818\times10^{5}$ & 0.418 & $1.069\times10^{-8}$ \\
$p_4$ & 9.925 & $1.082\times10^{-7}$ & $1.471\times10^{6}$ & 0.455 & $4.956\times10^{-9}$ \\
\hline\hline
\end{tabular}
\end{table}

\begin{figure}[t]
\includegraphics[width=\linewidth]{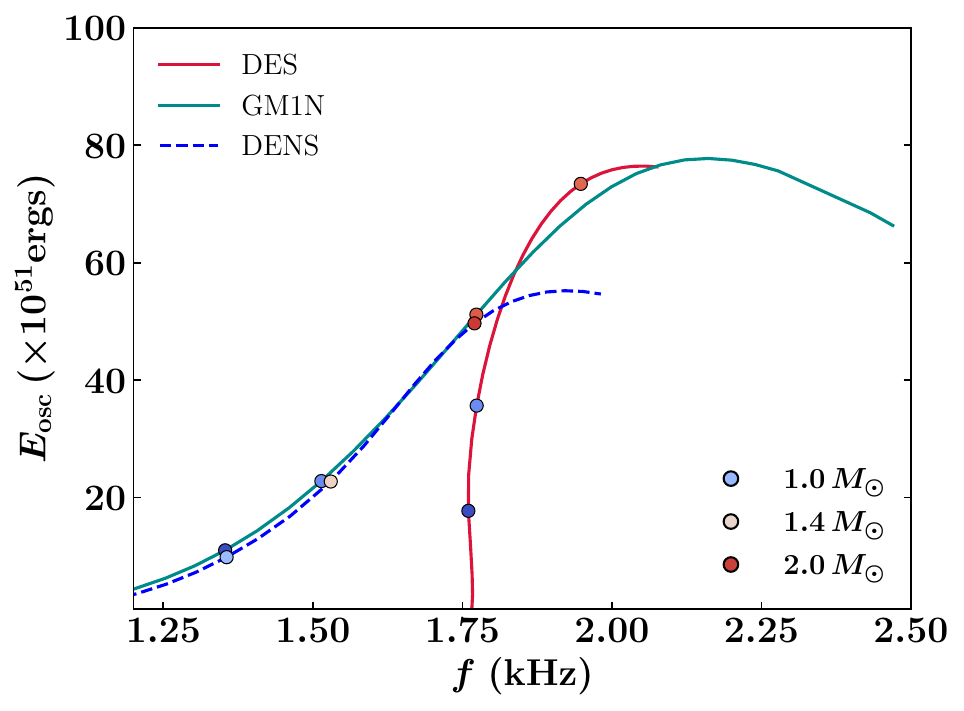}
\caption{The oscillation energy ($E_{\rm osc}$) as a function of $f$-mode frequency for the neutron star (GM1N), pure DE star (DES), and hybrid neutron star with a DE core (DENS).\label{fig:energy-f} }
\end{figure}

\begin{table}[h!]
\centering
\caption{Complex quasi-normal mode frequencies for the GM1 equation of state with
$\ell=2$ and central pressure $p_c=8.15\times10^{-5}\,\mathrm{km}^{-2}$.
The corresponding stellar model has radius $R=13.27\,\mathrm{km}$ and mass
$M=1.56\,M_{\odot}$.}
\label{tab:gm1n}
\small
\begin{tabular}{lccccc}
\hline\hline
Mode & Re($f$) & Im($f$) & $\tau$ & $\omega_rM$ & $\omega_iM$ \\
     & (kHz) & (kHz) & (ms) & & \\
\hline
$f$   & 1.661 & $6.824\times10^{-4}$ & 233.2 & 0.080 & $3.291\times10^{-5}$ \\
$p_1$ & 5.076 & $4.238\times10^{-5}$ & $3.756\times10^{3}$ & 0.245 & $2.044\times10^{-6}$ \\
$p_2$ & 7.830 & $1.049\times10^{-7}$ & $1.518\times10^{6}$ & 0.378 & $5.058\times10^{-9}$ \\
$p_3$ & 10.032 & $7.726\times10^{-8}$ & $2.060\times10^{6}$ & 0.484 & $3.726\times10^{-9}$ \\
\hline\hline
\end{tabular}
\end{table}
%%%%%%%%%%%%%%%%%%%%%%%%%%%%%%%%%%%%%%%%%%%%%%%%%%%%%%%%%%%%%%%%%%%%%%%%%%%%%%%%%%%%%%%%%%%
In Fig.~\ref{fig:energy-f}, we show the variation of the oscillation energy $E_{\rm osc}$ with $f$-mode frequency for three models: pure NS (GM1N), pure DE star (DES), and NS with a DE core (DENS). The oscillation energy corresponding to the $f$-mode frequency can be numerically obtained by integrating Eq.~\ref{eq:E_osc} from the center to the surface of the star. 
For all the three stellar models, the oscillation energy is found to increase with the $f$-mode frequency up to the vicinity of the maximum-mass configuration, beyond which it decreases. We find that at lower frequencies, the DENS configurations closely follow the GM1N sequence, indicating that the oscillation properties are dominated by NS matter. However, as the frequency increases, the DENS model gradually deviates from the GM1N curve, clearly depicting the influence of DE core. 
Interestingly, although the presence of DE matter considerably influences the $E_{\rm osc}$, the DENS curve does not converge towards that of DES. Instead, it closely follows the GM1N curve over most of the frequency range, indicating that the global oscillation properties are primarily governed by the NS matter. Thus, the DE core modifies the oscillation energy while largely preserving the $f$-mode characteristics of NSs, except in the high-frequency regime. 

%%%%%%%%%%%%%%%%%%%%%%%%%%%%%%%%%%%%%%%%%%%%%%%%%%%%%%%%%%%%%%%%%%%%%%%%%%%%%%%%%%%%%%%%%%%%%
\begin{figure}[h!]
\includegraphics[width=\linewidth]{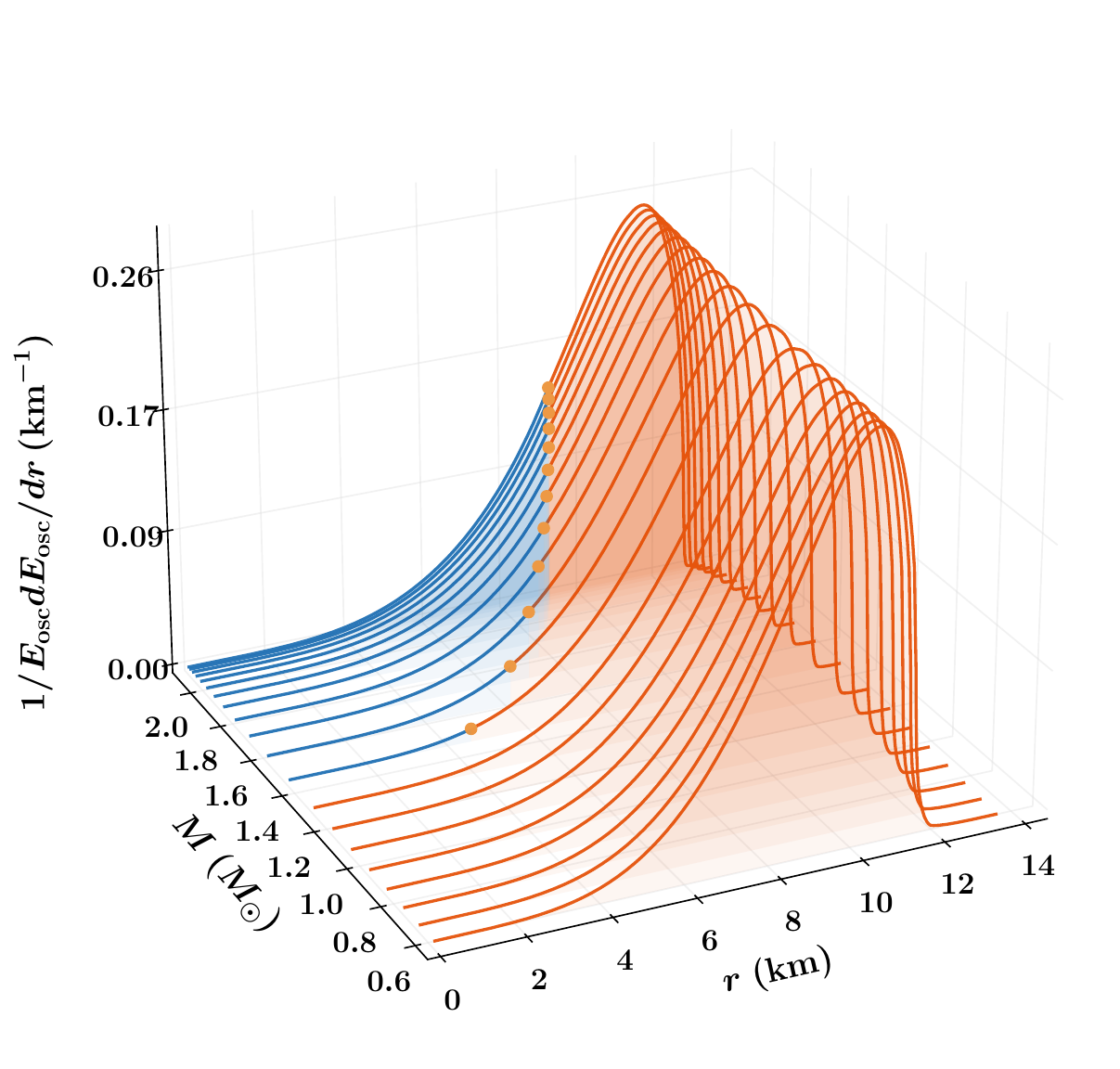}
\caption{The scaled gradient of oscillation energy $(dE_{\rm osc}/dr)/E_{\rm osc}$ as function of radius $r$ for a NS with DE core. The markers indicate the corresponding DE core boundaries. \label{fig:energy-profiles} }
\end{figure}
Next, we examine the scaled oscillation energy $(dE_{\rm osc}/dr)/E_{\rm osc}$ profile using the Eq.~\ref{eq:E_osc} for a hybrid DENS model with $A=0.2$ and $\alpha=1$. 
In Fig.~\ref{fig:energy-profiles},  
the red and blue colour correspond to the regions $r<r_t$ and $r>r_t$, respectively. 
Here, $r_t$ is determined by the transition density $\varepsilon_t=0.55\times10^{15}\,{\rm g\,cm^{-3}}$, and the markers indicate the corresponding location of the DE core boundary. 
For all mass profiles, we find that the scaled gradient of the oscillation energy initially increases from the stellar center, reaches a maximum at an intermediate radius, and then decreases rapidly toward the stellar surface. 
These profiles largely overlap radially, with only slight shifts observed in the higher-mass region. 
We find an increase in the transition radius ($r_t$) with an increase in stellar mass, indicating a greater radial extent of the DE component in more massive stars.
Further, we find that the maximum scaled oscillation energy lies around $r\simeq10.35-12.00$ km for the stellar masses considered are in the range $0.23-0.26$ km$^{-1}$. The corresponding maximum values of $dE_{\rm osc}/dr$ are found to increase with increasing stellar mass, resulting in a greater contribution of the oscillation energy in more massive stellar configurations. 
%%%%%%%%%%%%%%%%%%%%%%%%%%%%%%%%%%%%%%%%%%%%%%%%%%%%%%%%%%%%%%%%%%%%%%%%%%%%%%%%%%%%%%%%%%%

\begin{figure}[h]
\includegraphics[width=\linewidth]{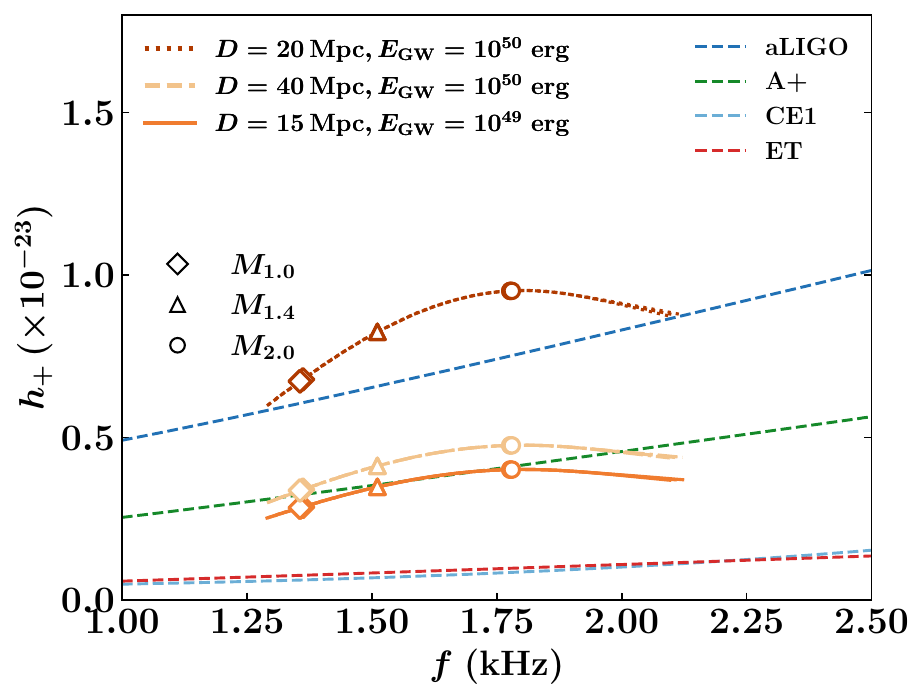}
\caption{Characteristic GW strain ($h_+$) as a function of the $f$-mode frequency for representative NS models with DE cores. The detector sensitivity curves of Advanced LIGO (aLIGO)~\cite{KAGRA:2013rdx}, A+~\cite{aplus}, Cosmic Explorer (CE1), and the Einstein Telescope (ET)~\cite{Hild:2010id} are shown for comparison. The diamond, triangle, and circle symbols correspond to the stellar models with masses of $1.0\,M_\odot$, $1.4\,M_\odot$, and $2.0\,M_\odot$, respectively. The predicted characteristic strain amplitudes are compared with the detector sensitivities to evaluate the detectability of the $f$-mode GW signals.\label{fig:strain}}
\end{figure}
In Fig.~\ref{fig:strain}, we plot the characteristic GW strain as a function of the $f$-mode frequency for hybrid DENS considered. We also show the sensitivity curves of Advanced LIGO, A+, Cosmic Explorer, and the Einstein Telescope. We find that the predicted strain amplitudes for our models lie in the range of ~$h_+ \sim (0.3-1.0)\times 10^{-23}$ for a frequency span of $~1.35-2.10$ kHz. As the stellar mass increases, both the oscillation frequency and the strain amplitude initially increase, whereas the strain decreases slightly at higher frequencies owing to its inverse dependence on the mode frequency. The increase in strain amplitude from the $M_{1.0}$ to $M_{2.0}$ configurations suggests that more compact stellar models are capable of radiating stronger $f$-mode GW signals.

%%%%%%%%%%%%%%%%%%%%%%%%%%%%%%%%%%%%%%%%%%%%%%%Summary&Conclusions%%%%%%%%%%%%%%%%%%%%%%%%%%%%%%%%%%%%%%%%%%%%%%%%%%%%%%%%%%%%%%%%%%%%%%%%%
\section{Summary and Conclusions}
\label{sec:summary}

In this work, we have presented the first study of non-radial $f$-mode oscillations of neutron stars (NSs)  containing a dark energy (DE) core, and investigated their static and dynamical properties within a fully relativistic framework. We considered a hybrid model composed of neutron matter and DE, with the latter described by a generalised Chaplygin gas EoS. We examined the effects of the DE parameters $A$ and $\alpha$, as well as the transition density $\varepsilon_t$ that determines the extent of the DE component within the stellar interior, on the stellar structure, tidal deformability, non-radial oscillation frequencies, damping times, oscillation energy, and the associated gravitational wave (GW) strain.

We first analysed the resulting EoSs and found that increasing either $A$ or $\alpha$ softens the hybrid EoS. The choice of transition density plays an important role in determining the extent to which the DE component influences the stellar properties. For the models obtained by varying $A$, the relatively high transition density confines the DE component predominantly to the inner core, whereas the lower transition density adopted for the $\alpha$-variation models allows DE matter to extend over a substantially larger fraction of the stellar core. All the models considered satisfy the causality condition, $c_s^2<1$, and the dynamical stability condition, $\Gamma>4/3$. The inclusion of DE reduces $\Gamma$ as the transition density is approached, no violation of the stability criterion is found for the parameter space explored here.

We find that the presence of a DE core systematically modifies the global stellar structure, leading to a reduction of the maximum mass relative to the corresponding pure NS sequence. The extent of these modifications is strongly controlled by the transition density. For the $A$-variation models, the higher transition density confines the deviations from the pure NS sequence mainly to the high-mass region. In contrast, the lower transition density adopted for the $\alpha$-variation models allows the DE component to affect the stellar structure over a substantially broader mass range. The resulting mass-radius sequences are consistent with observational data from massive pulsars and X-ray observations such as PSR J0030+045 and PSR
J0740+6620. This demonstrates that the location of the NS-DE interface is an important parameter in determining the macroscopic imprint of the DE core.

The non-radial $f$-mode spectrum provides a direct probe of these structural modifications. Employing the fully relativistic perturbative treatment, we then numerically computed the eigen frequencies of the $f$-mode oscillations and found that the inclusion of DE produces systematic changes in the oscillation frequencies and damping times. In particular, the range of frequencies spanned by the $\alpha$-variation models is narrower than that obtained by varying $A$, reflecting the different ways in which the two DE parameters modify the stellar structure. The corresponding changes in the damping times indicate that the presence of a DE core also modifies the GW emission properties of the $f$-mode. Further, we studied the correlation between the $f$-mode frequency and tidal deformability $\Lambda$ and found that the DE core configurations remain consistent with the observational constraints from GW170817 and GW190814, with their sequences approaching the pure NS relation at higher $f$-mode frequencies.

An interesting feature emerges from the radial distribution of the oscillation energy. Although the presence of a DE core smoothens the energy profile, it has only a negligible effect on the radial localisation of the mode energy. Moreover, for the DE core model considered, the normalized energy gradient exhibits an approximately universal behaviour over the mass range $1.0$ - $1.9,\,M_\odot$. This indicates that, despite the substantial modifications of the equilibrium structure produced by the DE core, the spatial character of the $f$-mode remains largely governed by the outer stellar layers.

Finally, using the calculated $f$-mode frequencies and damping times, we estimated the corresponding characteristic GW strain. The strain amplitude increases with stellar mass over the range $1.0$-$2.0,\,M_\odot$ within our adopted excitation model, with the resulting signals lying within the sensitivity range of future third-generation GW detectors. Taken together, our results show that a DE core can leave correlated imprints on the mass-radius relation, tidal deformability, $f$-mode spectrum, damping time, and GW strain, while preserving the overall spatial character of the oscillation mode. These results demonstrate that the presence and extent of a DE core in NSs can produce appreciable modifications in the global stellar properties and $f$-mode GW observables, while leaving the radial localisation of the oscillation energy largely unaffected.

\bibliography{ref}

\end{document}